\documentclass{article}

\usepackage{graphicx}
\usepackage{url}
\usepackage{eurosym}

\newif\ifpdf\ifx\pdfoutput\undefined\pdffalse\else\pdfoutput=1\pdftrue\fi

\ifpdf\fi

\begin{document}

\title{MIREX:\\MapReduce Information Retrieval Experiments
\thanks{Published as CTIT Technical Report TR-CTIT-10-15, ISSN 1381-3625, April 2010.}
}

\author{ 
Djoerd Hiemstra and Claudia Hauff\\
 University of Twente, The Netherlands\\
 \{d.hiemstra, c.hauff\}@cs.utwente.nl
}
\date{}
\maketitle

\begin{abstract}
{We propose to use MapReduce to quickly test new retrieval approaches on a cluster of machines by sequentially scanning all documents. We present a small case study in which we use a cluster of 15 low cost machines to search a web crawl of 0.5 billion pages showing that sequential scanning is a viable approach to running large-scale information retrieval experiments with little effort. 
The code is available to other researchers at: \url{http://mirex.sourceforge.net}
}
\end{abstract}

\section{Introduction}
A lot of research in the field of information retrieval aims at improving the {\em quality} of search results. Search quality might for instance be improved by new scoring functions, new indexing approaches, new query (re-)formulation approaches, etc. To make a scientific judgment of the quality of a new search approach, it is good practice to use so-called benchmark test collections, such as those provided by TREC \cite{TrecBook}. The following steps typically need to be taken: 
\begin{enumerate}
\item The researcher codes the new approach by adapting an experimental search system, such as Lemur \cite{Lemur}, PF/Tijah \cite{PFTijah}, or Terrier \cite{Terrier}; 
\item The researcher uses the system to create an inverted index on the documents from the test collection; 
\item The researcher puts the queries to the experimental search engine and gathers the top $X$ search results (a common value for TREC experiments is $X=1000$); 
\item The researcher compares the top $X$ to a golden standard by computing standard evaluation measures such as mean average precision.
\end{enumerate}
In our experience, Step 1, actually coding the new approach, takes by far the most effort and time when conducting an information retrieval experiment. Coding new retrieval approaches into existing search engines like Lemur, PF/Tijah and Terrier is a tedious job, even if the code is maintained by members of the same research team. It requires detailed knowledge of the existing code of the search engine, or at least, knowledge of the part of the code that needs to be adapted. Radical new approaches to information retrieval, i.e., approaches that need information that is not available from the search engines inverted index, require reimplementing part of the indexing functionality. Such radical new approaches are therefore not often evaluated, and most research is done by small changes to the system.

In his WSDM keynote lecture, Dean \cite{Dean09WSDM} describes how Map\-Reduce \cite{MapReduce} is used at Google for experimental evaluations. New ranking ideas are tested off-line on human rated query sets similar to the queries from TREC\@. Running such off-line tests has to be easy for the researchers at Google, possibly at the expense of the efficiency of the prototype. So, it is okay if it takes hours to run for instance 10,000 queries, as long as the experimental infrastructure allows for fast and easy coding of new approaches. 
A similar experimental setup was followed by Microsoft at TREC 2009: Craswell et al.\ \cite{Craswell09} use DryadLINQ \cite{DryadLINQ} on a cluster of 240 machines to run web search experiments. Their setup also sequentially scans all document representations, providing a flexible environment for a wide range of experiments. The researchers plan to do many more to discover its benefits and limitations.

The work at Google and Microsoft shows that sequential scanning over large document collections is a viable approach to experimental information retrieval. Some of the advantages are: 
\begin{enumerate}
\item Researchers spend less time on coding and debugging new experimental retrieval approaches;
\item It is easy to include new information in the ranking algorithm, even if that information would not normally be included in the search engine's inverted index;
\item Researchers are able to oversee all or most of the code used in the experiment;
\item Large-scale experiments can be done in reasonable time.
\end{enumerate}

\noindent
We show that indeed sequential scanning is a viable experimental tool, even if only a few machines are available. In Section \ref{sec:approach} we describe the MapReduce search system. Sections \ref{sec:results} and \ref{sec:conclusion} contain experimental results and concluding remarks.

\section{Sequential Search in MapReduce}
\label{sec:approach}
MapReduce is a framework for batch processing of large data sets on clusters of commodity machines \cite{MapReduce}. Users of the framework specify a {\em mapper} function that processes a key/value pair to generate a set of intermediate key/value pairs, and a {\em reducer} function that processes intermediate values associated with the same intermediate key. The pseudo code in Figure \ref{tab:pseudocode} outlines our sequential search implementation. The implementation does a single scan of the documents, processing all queries in parallel. 

\begin{figure}[htb]
\begin{center}  \begin{sffamily}
\begin{tabular}{|l|}
\hline
{\tt ~}\\
{\tt ~mapper (DocId, DocText) = }\\
{\tt ~~~FOREACH (QueryID, QueryText) IN Queries}\\
{\tt ~~~~~Score = experimental\_score(QueryText, DocText)~}\\
{\tt ~~~~~IF (Score > 0)}\\ 
{\tt ~~~~~THEN OUTPUT(QueryId, (DocId, Score))}\\
{\tt ~}\\
{\tt ~reducer (QueryId, DocIdScorePairs) = }\\
{\tt ~~~RankedList = ARRAY[1000]}\\
{\tt ~~~FOREACH (DocId, Score) IN DocIdScorePairs}\\
{\tt ~~~~~IF (NOT filled(RankedList) OR }\\
{\tt ~~~~~~~Score > smallest\_score(RankedList))}\\
{\tt ~~~~~THEN ranked\_insert(RankedList, (DocId, Score))~}\\
{\tt ~~~FOREACH (DocId, Score) IN RankedList}\\
{\tt ~~~~~OUTPUT(QueryId, DocId, Score)}\\
{\tt\tiny ~}\\
\hline
\end{tabular}
\caption{Pseudo code for linear search}
\label{tab:pseudocode}
\vspace{-4mm}
\end{sffamily}\end{center}
\end{figure}

The {\em mapper} function takes pairs of {\em document identifier} and {\em document text} {\tt (DocId, DocText)}. For each pair, it runs all benchmark queries and outputs for each matching query the {\em query identifier} as key, and the pair {\em document identifier} and {\em score} as value. In the code, {\tt Queries} is a global constant per experiment. The MapReduce framework runs the mappers in parallel on each machine in the cluster. When the map step finishes, the framework groups the intermediate output per key, i.e., per {\tt QueryId}. The {\em reducer} function then simply takes the top 1000 results for each query identifier, and outputs those as the final result. The reducer fucntion is also applied locally on each machine (that is, the reducer is also used as a  {\em combiner} \cite{MapReduce}), making sure that at most 1000 results have to be sent between machines after the map phase finishes.

\section{Case Study: ClueWeb09}
\label{sec:results}
\noindent
The ClueWeb09 test collection consists of 1 billion web pages in ten languages, collected in January and February 2009. The dataset is used by several tracks of the TREC conference \cite{TrecBook}. We used the English pages from the collection, about 0.5 billion pages equalling 12.5 TB (2.5 TB compressed). The size of the ClueWeb09 collection cannot be handled by a single machine,  unless one is willing to buy special hardware. We ran our experiments on a small cluster of 15 machines; each machine costs about {\euro}\,1000\@. The cluster runs Hadoop version 0.19.2 out of the box \cite{HadoopBook}.

\subsection{Time to code the experiment}
After gaining some experience with Hadoop by having M.Sc.\ students doing practical assignments, we wrote the code for sequential search, and for anchor text extraction in less than a day. 
Table \ref{tab:size} gives some idea of the size of the source code compared to that of experimental search systems. Note that this is by no means a fair comparison: The existing systems are general purpose information retrieval systems including a lot of functionality, whereas the linear search system only knows a single trick. Still, in order to adapt the systems below, one at least has to figure out what code to adapt.

\begin{table}[htb]
\begin{center}  \begin{normalsize} \begin{sffamily}
\begin{tabular}{l|r|r|r}
Code base                  & ~~\#files & \#lines  & size (kb)\\
\hline
MapReduce anchors \& search~~&     2   &       350 &     13 \\
Terrier 2.2.1              &   300   &    59,000 &  2,000 \\
MonetDB/PF/Tijah 0.32.2    &   920   & ~~1,393,000 & 40,600 \\ 
Lemur/Indri 4.11           & 1,210   &   540,000 & 19,500 \\ 
\end{tabular}
\caption{Size of code base per system}
\label{tab:size} 
\end{sffamily}\end{normalsize}\end{center}
\end{table}

\subsection{Time to run the experiment}
Anchor text extraction on all English documents of Clue\-Web09 takes about 11 hours on our cluster. The anchor text representation contains text for about 87~\% of the documents, about 400 GB in total. A subsequent TREC run using 50 queries on the anchor text representation takes less than 30 minutes. 
Our linear search system implements a fairly simple language model with a length prior without stemming or stop words. It achieves expected precision at 5, 10 and 20 documents retrieved of respectively  0.42,  0.39, and 0.35 (MTC method), similar to the best runs at TREC 2009~\cite{TREC09Overview}. 

\begin{figure}[htb]
\begin{center}  \begin{normalsize} \begin{sffamily}
\includegraphics[width=10.6cm]{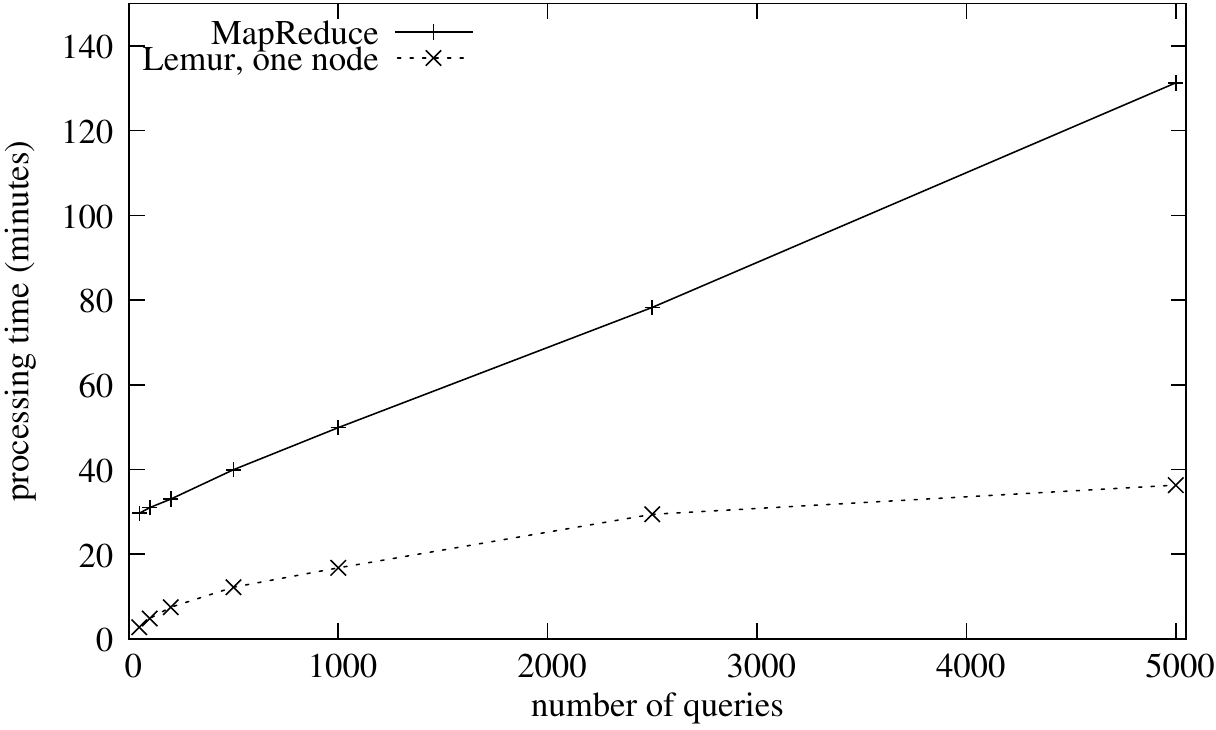}
\caption{Processing time for query set sizes}
\label{fig:time}
\end{sffamily}\end{normalsize}\end{center}
\end{figure}

Figure \ref{fig:time} shows how the system scales when processing up to 5,000 queries, using random sets of queries from the TREC 2009 Million Query track. Reported times are full Hadoop job times including job setup and job cleanup averaged over three trials.
Processing time increases only slightly if more queries are processed. Whereas the average processing time per query is about 35 seconds per query for 50 queries, it goes down to only 1.6 second per query for 5,000 queries. 
For comparison, the graph shows the performance of ``Lemur-one-node'', i.e., Lemur version 4.11 running on {\em one fourteenth} of the anchor text representation on a single machine. A distributed version of Lemur searching the full full anchor text representation would not do faster: It would be as fast as the slowest node, it would need to send results from each node to the master, and to merge the results. Lemur-one-node takes 3.3 seconds per query on average for 50 queries, and 0.44 seconds on average for 5,000 queries. The processing times for Lemur were measured after flushing the file system cache. Although Lemur cannot process queries in parallel, the system's performance benefits a lot from receiving a lot of queries. Lemur's performance scales sublinearly because it caches intermediate results.  Still, at 5,000 queries Lemur-one-node is only 3.6 times faster than the MapReduce system. For experiments at this scale, the benefits of the full, distributed Lemur are probably negligible. 

\subsection{Related work}
The idea to use sequential scanning of documents to research new retrieval approaches is certainly not new: We know of at least one researcher who used sequential scanning over ten years ago for his thesis \cite{Hiemstra01}. Without high-level programming paradigms like MapReduce, however, efficiently implementing sequential scanning is not a trivial task, and without a cluster of machines the approach does not scale to large collections. 

Lin \cite{Lin09} used Hadoop MapReduce for computing pairwise document similarities. Our implementation resembles Lin's brute force algorithm that also scans document representations linearly. Our approach is simpeler because our preprocessing step does not divide the collection into blocks, nor does it compute document vectors.

\section{Conclusion}
\label{sec:conclusion}

\noindent
A faster turnaround of the experimental cycle can be achieved by making  coding of experimental systems easier. Faster coding means one is able to do more experiments, and more experiments means more improvement of retrieval performance. 
We implemented a full experimental retrieval system with little effort using Hadoop MapReduce.
Using 15 machines to search a web crawl of 0.5 billion pages, the proposed MapReduce approach is less than 10 times slower than a single node of a distributed inverted index search system on a set of 50 queries. If more queries are processed per experiment, the processing times of the two systems get even more close. 
The code used in our experiment is open source and available to other researchers at: \url{http://mirex.sourceforge.net}

\section*{Acknowledgments}
Many thanks to Sietse ten Hoeve, Guido van der Zanden, and Michael Meijer for early implementations of the system. The research was partly funded by the Netherlands Organization for Scientific Research, NWO, grant 639.022.809\@. We are grateful to Yahoo Research, Barcelona, for sponsoring our cluster.


\end{document}